%% file: main.tex
\documentclass[preprint2]{aastex62}

\usepackage{natbib}
\usepackage{graphicx}
\usepackage{amsmath} 
\usepackage{mathtools}
\usepackage{newtxtext,newtxmath}

\begin{document}
\title{Kepler-80 Revisited: Assessing the Participation of a Newly Discovered Planet in the Resonant Chain}


\author[0000-0002-7992-469X]{D.~Weisserman}
\affiliation{Department of Astronomy, University of Michigan, Ann Arbor, MI 48109, USA}
\author[0000-0002-7733-4522]{J.~C.~Becker}
\affiliation{Division of Geological and Planetary Sciences, California Institute of Technology, Pasadena, CA 91125, USA}
\author[ 0000-0001-7246-5438]{A.~Vanderburg}
\affiliation{Department of Physics and Kavli Institute for Astrophysics and Space Research, Massachusetts Institute of Technology, Cambridge, MA 02139, USA}

\begin{abstract}
In this paper, we consider the chain of resonances in the Kepler-80 system and evaluate the impact that the additional member of the resonant chain discovered by \citet{Shallue2018} has on the dynamics of the system and the physical parameters that can be recovered by a fit to the transit timing variations (TTVs). Ultimately, {we calculate the mass of Kepler-80 g to be $0.8 \pm 0.3 M_\oplus$ when assuming all planets have zero eccentricity, and $ 1.0 \pm 0.3 \ M_{\oplus}$ when relaxing that assumption. }
We show that the outer five planets are in successive three-body {mean-motion resonances (MMRs)}. We assess the current state of two-body {MMRs} in the system and find that the planets do not appear to be in two-body {MMRs}. We find that while the existence of the additional member of the resonant chain does not significantly alter the character of the Kepler-80 three-body MMRs, it can alter the physical parameters derived from the TTVs, suggesting caution should be applied when drawing conclusions from TTVs for potentially incomplete systems. {We also compare our results to those of \citet{MacDonald2021}, who perform a similar analysis on the same system with a different method. Although the results of this work and \citet{MacDonald2021} show that different fit methodologies and underlying assumptions can result in different measured orbital parameters, the most secure conclusion is that which holds true across all lines of analysis: Kepler-80 contains a chain of planets in three-body MMRs but not in two-body MMRs.}
\end{abstract}

\section{Introduction} \label{sec:intro}
The existence of {chains of exoplanets residing in mean-motion resonances} provides a unique constraint on their host systems' evolutionary histories. The most common explanation for a present-day {mean-motion} orbital resonance is that the resonant configuration assembled via past disk-driven migration \citep{Terquem2007, Raymond2008, Rein2012, Batygin2015}. For a small, non-gap-opening planet residing in a protoplanetary disk, asymmetry in torques from the disk material interior and exterior to the planet can lead to planetary migration in semi-major axis \citep{Goldreich1979, Ward1997, Tanaka2002}. 
Generally, this migration will be inwards for small planets \citep{Tanaka2002, DAngelo2010, Paardekooper2010, Bitsch2014VII, Bitsch2014VIII}, except {near the exterior edges of disk cavities} \citep{Masset2006}. 
In the case where migration rates for adjacent planets allow convergent migration, the eventual result of this process can be the emergence of mean motion resonances \citep{Snellgrove2001, Papaloizou2005}. In high-multiplicity systems, this migration process can result in chains of resonant planets and multi-body resonances \citep{Cossou2013}, {and the specific orbital properties of the involved planets can be used to infer their systems' formation and migration histories \citep{Lee2002, Tamayo2017}. For this reason, to understand the origin of resonant chains, it is important to precisely measure the orbital properties of resonant planets.}

The observational data set has presented some intriguing chains of resonant planets {for which the orbital properties can be measured}. 
Some notable examples of multi-resonant systems include TRAPPIST-1 \citep{Gillon2017, Luger2017}, Gliese-876 \citep{Marcy2001, Lee2002}, {TOI-178 \citep{Leleu2021},} K2-138 \citep{Christiansen2018}, Kepler-223 \citep{Mills2016}, and Kepler-80 {\citep{Ford2011, MacDonald2016}}. 
However, one major factor preventing secure conclusions from discovered exoplanet systems is the likely ubiquitous presence of undiscovered planets in even well-studied systems. 
There have been many examples over the last decade of planet discovery where a multi-planet system is found to host additional planets{, and more will continue to be discovered even in systems that appear well-characterized.}
These planets might have been missed the first time because they were non-transiting \citep[as in Kepler-20,][]{Buchhave2016}, because their detections had too low of a signal to noise ratio to be detected {in the initial analysis of the} data \citep[as in Kepler-80 and Kepler-90,][]{Shallue2018}, or because of the timing of their transit events compared to the baselines of observation \citep{Santerne2019}. 
The presence of additional planets often necessitates the assessment of whether earlier dynamical analyses still hold with the additional perturber's presence in the system. 

In this paper, we evaluate the Kepler-80 planetary system in the context of its new planetary companion Kepler-80 g, reported in \citet{Shallue2018}. {We used a center-time-of-transit fit to constrain the mass of this additional planet, then assessed its participation in the resonant chain. A similar analysis on Kepler-80 g has been published in \citep{MacDonald2021}, which uses a photodynamical fit to the same data used in this work\footnote{We note that the work in this manuscript and \citet{MacDonald2021} was begun concurrently and separately, with the latter being published as this manuscript was first submitted to the journal.}. These two methods obtained very different results for the mass of Kepler-80 g. In this work, we discuss the two methods and their solutions with goals of (1) explaining the methodological differences that lead to the discrepancy, and (2) bringing attention to the ways that TTV fit solutions can misrepresent the state of a system.}
In Section \ref{sec:ttvs}, we explore the effect the additional planet has on the solution of a transit timing variation analysis aimed at recovering the planetary masses {and discuss how difference in methodology between this work and \citep{MacDonald2021} resulted in two different solutions for the planetary masses}.
In Section \ref{sec:resonance}, we consider the resonant state of the system, and whether the presence of the additional planet changes the resonant state. {We also compare the results of this work to previous results on the Kepler-80 system.}
In Section \ref{sec:disc}, we provide a discussion of the lessons learned from this Kepler-80 test case, and how future analysis of multi-planet systems can account for the presence of unseen planets. 

\section{{System Parameters from} Transit Timing Variations}
\label{sec:ttvs}

Transit-timing variations (TTVs) provide a method of {constraining} exoplanets' orbital parameters by measuring planets' deviations from purely Keplerian orbits.
Planet-planet interactions will lead planetary transit events to occur earlier or later than expected, and the amplitude and periodicity of these deviations allows for the detection of additional unseen planets \citep{Holman2005, Nesvorny2012, Nesvorny2013} or constraints on planetary masses \citep{Agol2005, Lithwick2012, Deck2015chop}. 
Orbital resonances can boost planet-planet interactions, resulting in larger TTV amplitudes. 
Because of the boost in TTV signal for planets near resonance, the TTV method has been most effectively used to constrain masses for planets orbiting {in or near} resonant configurations \citep{Xie2014}. 

{A TTV solution for the Kepler-80 system was first reported in \citet{MacDonald2016}. Since then, an additional planet was discovered in the Kepler-80 system in \citet{Shallue2018}. \citet{MacDonald2021} then performed a photodynamical fit to constrain the orbital parameters of five of the known planets in the system. In this work, we performed a center-time-of-transit fit to constrain the orbital parameters of the same planets that were considered in \citet{MacDonald2021}.}

{In addition to constraining the planetary orbital parameters using a different method than was used in \citet{MacDonald2021}, our secondary goal in this work is to ascertain the differences in results that can be obtained under a series of different assumptions using a single fit method. To do this, we ran four center-time-of-transit fits with varying assumptions. These fits vary across two dimensions: first, whether Kepler-80 g (the new system component discovered in \citet{Shallue2018}) is included in the model; second, whether planetary eccentricities are allowed to vary.}

\subsection{Comparing Orbital Solutions for $e=0$ fits}
\label{sec:nullecc}

{We first ran two fits where we assumed the planetary eccentricities are zero: one fit with the system as it was known prior to the discovery of Kepler-80 g, and a second fit including Kepler-80 g. The purpose of these two fits are to assess how the addition of the extra planet to the system changes the orbital parameters derived from the measured TTVs under the assumption that the orbits are all circular}.

We performed our Markov-Chain Monte Carlo (MCMC) fits using the affine-invariant sampler implemented in the \texttt{emcee} package \citep{emcee1, emcee2, emcee3} and the TTV model implemented in {TTVFast \citep{Deck2014, Deck2014soft}.}
We used the transit times from \citet{MacDonald2016} and ran two initial MCMC fits (one including Kepler-80 g in the model, and one excluding it) using TTVFast to generate orbital solutions that match the observed mid-transit times. 

For both these fits, all planetary eccentricities had a delta function prior at 0; all planetary masses were given a uniform prior between 0 and 100 $M_{\oplus}$; {the stellar mass was given a Gaussian prior centered at 0.73 $M_{\odot}$ with a standard deviation of 0.03 $M_{\odot}$}. {The inclinations and longitudes of ascending nodes of all planets were fixed to 90° and 0° respectively.} The priors on orbital periods of all planets {included in the fit} were uniform priors ranging from 80\% and 120\% of the best-fit values given in \citet{MacDonald2016}. {As part of the purpose of this work is to assess the effect of the underlying assumptions, these priors were kept intentionally uninformative so as to not bias the results towards a previously computed solution.}
Overlapping transits were not used in the fit, but otherwise all events from {the short-cadence Kepler data} for Kepler-80 b, c, d, and e were included. Kepler-80 f, the ultra-short period planet in the system, was not modeled because it is dynamically decoupled from the rest of the system \citep{MacDonald2016}. 
{The first \texttt{emcee} fit was run without Kepler-80 g. This first fit was run with 12 model free parameters and 128 walkers for 400000 steps, which yielded an acceptance fraction of 36\%. The chain length was over 50 times the autocorrelation time, indicating that the chains are well-mixed and the solution converged. 10\% of the chain was removed as burn-in. The results of this fit are} shown in Table \ref{bigtable}.

The second fit was {run similarly to the previous fit, but} with an additional planet added to the model, to account for the existence of Kepler-80 g \citep{Shallue2018}. 
Transit {times} for Kepler-80 g were not {modeled directly} due to their low signal-to-noise and lack of precise individual transit times. Instead, the fit {likelihood function} was adapted to match the data type available. \citet{Shallue2018} computed the best-fit orbital period and time of transit based on the data available from \emph{Kepler} but without accounting for TTVs, meaning that the values reported in that work are the time of the first transit seen in \emph{Kepler} and the mean orbital period over the \emph{Kepler} baseline. Depending on what segment of the TTV super-period was observed during the \emph{Kepler} baseline, the observed mean orbital period may deviate from the true orbital period. As such, in this second fit, while we fit for the same parameters as done previously (the mass, orbital period, and mean anomaly) for the inner planets, for Kepler-80 g we fit for the transit time of the first event in Kepler and its {mean} orbital period {over the Kepler baseline}. 
A Gaussian prior on the average period of Kepler-80 g computed during the Kepler baseline was used, with a central value of 14.64558 days and standard deviation 0.00012 days, in accordance with the values given in \citet{Shallue2018}. The center time of transit of the first event seen in the Kepler data for Kepler-80 g was given a Gaussian prior with mean 2455658.6073 BJD and standard deviation 0.0037 days, again in accordance with \citet{Shallue2018}.
{In this second \texttt{emcee} fit, we had 15 model free parameters and ran a fit with 128 walkers for 400000 steps, which yielded an acceptance fraction of 34\%. The chain length was over 50 times the autocorrelation time, indicating that the chains are well-mixed. 10\% of the chain was removed as burn-in. }
The results of the fit {are} shown in Table \ref{bigtable}. 


{Table \ref{bigtable} summarizes the difference between the fits including and excluding Kepler-80 g.} The main difference between the TTV fit with and without Kepler-80 g occurs in the masses of the planets, {which are $1\sigma$ consistent between the two cases but have different central values. }


 \begin{figure}
    \centering
    \includegraphics[width = 3.0in]{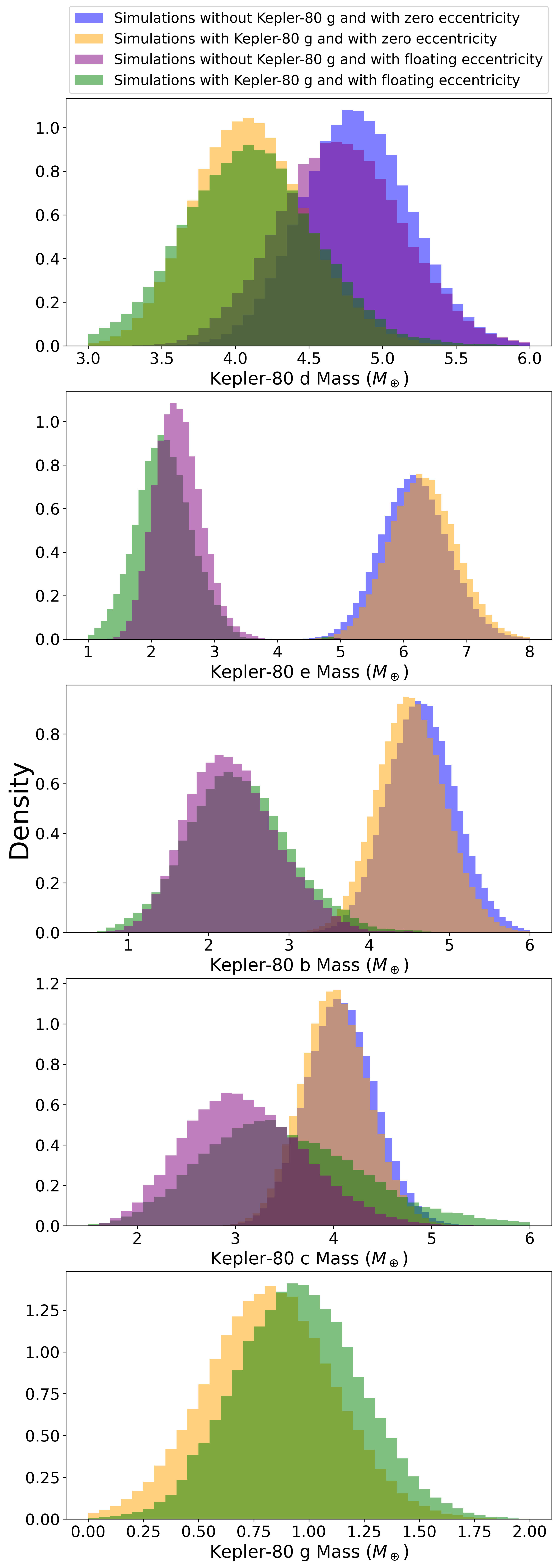}
    \caption{These plots show the distribution of possible values for the masses of each planet {under the four permutations of our fit assumptions.} }
    \label{fig:histogram_per_mass}
\end{figure}

\subsection{{Orbital Solutions for Fits with no Priors on Eccentricity}}
\label{sec:nonzeroecc}
In the fits considered in the previous section, the planetary eccentricities were set to 0. 
This assumption forces the TTV amplitude to be determined solely by the relative planetary masses. 
In reality, the TTV amplitude is a function of both the object masses and also the {relative} ($e \cos{\omega}$, $e \sin{\omega}$) of the planetary orbits. 
From our initial fits with zero eccentricity, we found that adding the additional planet to the system {does alter} the measured masses of {all the planets in the system, but not to statistically significant levels.}
However, once planets are allowed to have non-zero ($e \cos{\omega}$, $e \sin{\omega}$) values, the interactions between planets becomes more complex since planetary mass and eccentricity are degenerate parameters. 
With a sufficiently large quantity of data, this degeneracy can be broken with use of the synodic chopping signal \citep{Deck2015chop}, but this is not always possible since the chopping signal requires a higher level of precision on the transit times than is needed to see the bulk sinusoidal signal. 
To attempt to evaluate the full system state, we next expanded our {free parameters} to include the $\sqrt{e} \cos{\omega}$ and $\sqrt{e} \sin{\omega}$ for each planet, for a total of 25 fit parameters. We altered the fit to directly fit time of transit for all planets, since our previous fit parameter, the mean anomaly, depends on the argument of pericenter $\omega$ which is now allowed to float. 

{We ran a second \texttt{emcee} simulation with 25 model free parameters and 96 walkers for 600000 steps, which yielded an acceptance fraction of 22\%. This fit inherited the priors used in the previous fit except for those on planetary eccentricities. We placed no priors on the orbital eccentricities $e$ or arguments of pericenter $\omega$.} The chain length was over 50 times the autocorrelation time, indicating that the chains are well-mixed {and the fit converged}. The results of this fit are shown in Table \ref{bigtable}. With the relaxed assumption on the planetary eccentricities, the planetary masses are again slightly altered compared to the null-eccentricity fit given in Table \ref{bigtable}. The fit results suggest that {while Kepler-80 e, c, and g may have the largest eccentricities in this system, the system is nonetheless rather dynamically cold, with quite small eccentricities even when eccentricity is allowed to float.}

The {central value of the fitted mass distribution} of Kepler-80 g is slightly smaller ($ 0.8 ^{+0.8 }_{-0.6 } \ M_{\oplus}$) than the value computed in the null eccentricity fit (1.1 $\pm 0.4\ R_{\oplus}$), although the values are consistent within errors.


\subsection{{Comparison to the Results of \citet{MacDonald2021}}}
{As described above, the orbital constraints presented in Table \ref{bigtable} all use \texttt{TTVFast} to compute a model of center-time-of-transits and then compare those computed times to the times derived from the transit data. }
{As illustrated in Figure \ref{fig:histogram_per_mass}, the fit assumptions can significantly alter the derived best-fit masses, with cases where the planetary eccentricities are set to $e=0$ resulting in higher measured masses to such an extent that the solutions for the $e=0$ and floating eccentricity cases have inconsistent derived values for some parameters of some planets. }
{In contrast, including Kepler-80 g in the fit changes the best-fit value of the mass, but not so much that the solutions are inconsistent.}

{We can also make a comparison to the results of \citet{MacDonald2021}.
While, in this publication, we performed MCMC fits on the TTVs directly, \citet{MacDonald2021} fits the light curve of the star directly, which is a method called a photodynamical fit \citep[see also]{Barros2015}. This fit style fits a different set of parameters than used in this work. Notably, 
\citet{MacDonald2021} obtained significantly different planetary masses than those found in any of our fits ($5.95_{-0.60}^{+0.65} M_\oplus$, $2.97_{-0.65}^{+0.76} M_\oplus$, $3.50_{-0.57}^{+0.63} M_\oplus$, $3.49_{-0.57}^{+0.63} M_\oplus$, and $0.065_{-0.038}^{+0.044} M_\oplus$, respectively.) Among the fits produced as a result of this work, the floating eccentricity fit including Kepler-80 g (bottom rows of Table \ref{bigtable}) is the most representative of the known state of the system. The measured masses from that fit are generally $1 \sigma$ inconsistent with the results of \citet{MacDonald2021}. Most notably, the measured mass of Kepler-80 g is very different between the results of \citet{MacDonald2021} and our results here: the mass measured in \citet{MacDonald2021} was $0.065_{-0.038}^{+0.044} M_\oplus$, while the mass measured in this work was $0.8 \pm 0.4$ $M_\oplus$. The discrepancy is important to note because the mass of the outer planet have very different implications planetary migration rates, as well as the planetary composition.}

{This discrepancy can be explained by considering the fit methodology that was used in each paper. Photodynamical fits, as used in \citet{MacDonald2021}, generate a synthetic light curve and the optimization routine is performed on that light curve, rather than on the derived center times of transit as was done in this work. Photodynamical fits utilize one major additional constraint that is not captured in our model: the duration of transit events. The transit duration provides an additional constraint on $e \sin{(\omega)}$. However, the larger number of free parameters in the photodynamical model \citep[][uses 44 free parameters]{MacDonald2021} allows for the possibility of over-fitting for data with a low signal-to-noise ratio (SNR). 
In contrast, the center time of transit fit (used in this work) does not include any information about the transit durations, but as a result is not biased in the same way that the photodynamical fits are by low SNR data. Similarly, we chose our fit method to avoid directly fitting for the transit events of Kepler-80 g due to the low SNR of its individual transit events.}

{The fact that the results of our $e=0$ fits are inconsistent with the results of our floating eccentricity fits, which in turn are inconsistent with the results of \citet{MacDonald2021}, demonstrate the importance of careful assumptions for TTV fits and also the importance of careful interpretation of those results. In \citet{MacDonald2021}, the author state that conclusions should not be drawn from their measured mass of Kepler 80 g due to the issues of overfitting and low SNR data described above. In particular, their measured value of $e_g = 0.13$ is significantly higher than the values derived by the fits in this work, and this value could have been biased upwards due to poor SNR constraints on the transit durations of this planet. 
While the precision on our measured mass of Kepler-80 g in particular is significantly worse than the precision in \citet{MacDonald2021}, our best fit value is more physically realistic as our solution is not biased in the same way by low SNR data.}

{Future studies that use the parameters measured in any of the Kepler-80 fits \citep[in this work or in][]{MacDonald2021} to make implications about the formation and migration history of the system must account for the uncertainty. Additional data on the Kepler-80 system may help to ameliorate these difficulties.}
 

\section{{} Resonance}
\label{sec:resonance}

{Mean-motion orbital resonances occur when orbiting bodies exert regular gravitational influences on each other as a result of the bodies' periods being related by small-integer ratios of each other.} Three-body {mean-motion} resonances are a special type of resonance, which occurs when a configuration of three planets has a librating resonance angle given by:
\begin{equation}
\phi = p\lambda_1 - (p+q)\lambda_2 + q\lambda_3
\label{eq:threebodyres}
\end{equation}
where $\lambda = \Omega + \omega + M$ is the standard mean longitude of the planet, {and $\Omega$, $\omega$, and $M$ are the longitude of the ascending node, argument of periapsis, and mean anomaly of the planet, respectively}. Such a resonant configuration causes repeating conjunctions of planetary positions, subsequently causing periodic planet-planet perturbations. {The consistency of these periodic perturbations form a cycle that causes the result of Equation \ref{eq:threebodyres} to oscillate with some amplitude. This is called a libration, and the amplitude of the oscillations of Equation \ref{eq:threebodyres} is referred to as a libration amplitude.} 
When the angle defined by Equation \ref{eq:threebodyres} circulates between 0 and 360 degrees, then the planets are not in {a three-body mean-motion} resonance. 

The existence of {MMRs} can increase the {long-term} stability of the planetary system \citep[ex:][]{Tamayo2017, Obertas2017, Pichierri2018} as compared to a non-resonant configuration, although overlap of {MMRs} can also cause chaotic dynamics. In exoplanets, {MMRs} are taken as evidence of a disk-driven migratory history \citep{Rein2012conv, Paardekooper2013}, as planets will naturally fall into resonances as they convergently migrate inwards \citep{Cossou2013}. 

The Kepler-80 system, being {apparently} in a rare true multi-planet chain resonance, provides a strong test of convergent migration. Although multi-planet resonant chains are not a universal outcome of convergent migration \citep{Ketchum2011, Quillen2013, Pan2017, Hands2018, Pichierri2018}, systems where true resonance does exist provide unique insights to the dynamical histories of their planets \citep[ex:][]{Lee2002, MacDonald2018, Morrison2020, Goldberg2021}. For that reason, it is important to characterize the resonant state in multi-resonant systems exactly and to understand the robustness of that solution against the possibility of additional unseen planets or other perturbing effects. 

\subsection{Three body resonance in the presence of Kepler-80 g}
\label{sec:31}
One of the main results of the original analysis of the Kepler-80 system was the existence of a four planet resonant chain \citep{MacDonald2016}, a particularly unique architecture among the exoplanet sample. Since the presence of a true orbital mean-motion resonance depends rather sensitively on the orbital parameters of the involved planets, to assess whether a system is in an MMR it is important to construct the most accurate, complete orbital fit. {\citet{MacDonald2016} initially confirmed a four-planet resonant chain in Kepler-80. Following the discovery of Kepler-80 g \citep{Shallue2018}, \citet{MacDonald2021} found that the new planet likely does not participate in the resonant chain. However, Kepler-80 g's participation in the resonant chain depends sensitively on its measured mass, $e$, and $\omega$. As discussed in the previous section, our measured parameters for Kepler-80 g are significantly different from those presented in \citet{MacDonald2021}. As a result, in this section we reassess Kepler-80 g's participation in the resonant chain using the planetary parameters derived in this work.}

To determine the resonant state of the Kepler-80 system, we ran a suite of numerical simulations and evaluated the resonant behavior of each of the three potential three-body {MMRs}: Kepler-80 d, e, and b; Kepler-80 e, b, and c; and Kepler-80 b, c, and g. 
\citet{MacDonald2016} has shown that Kepler-80 c, b, and e are in tight-locking three-body {MMRs} (of equation $\phi = 2\lambda_c - 3\lambda_b + \lambda_e$), as are Kepler-80 d, e, and b (of equation $\phi = 2\lambda_d - 5\lambda_e + 3\lambda_b$.) 
Because of that, any new three-body {MMRs} would necessarily include Kepler-80 g. {We find that the new resonant angle is composed} of Kepler-80 g, c, and b ($p, q = 1$, with an equation given by $\phi = \lambda_g - 2\lambda_c + \lambda_b$.) 

\begin{figure*}
    \centering
    \includegraphics[width=15cm]{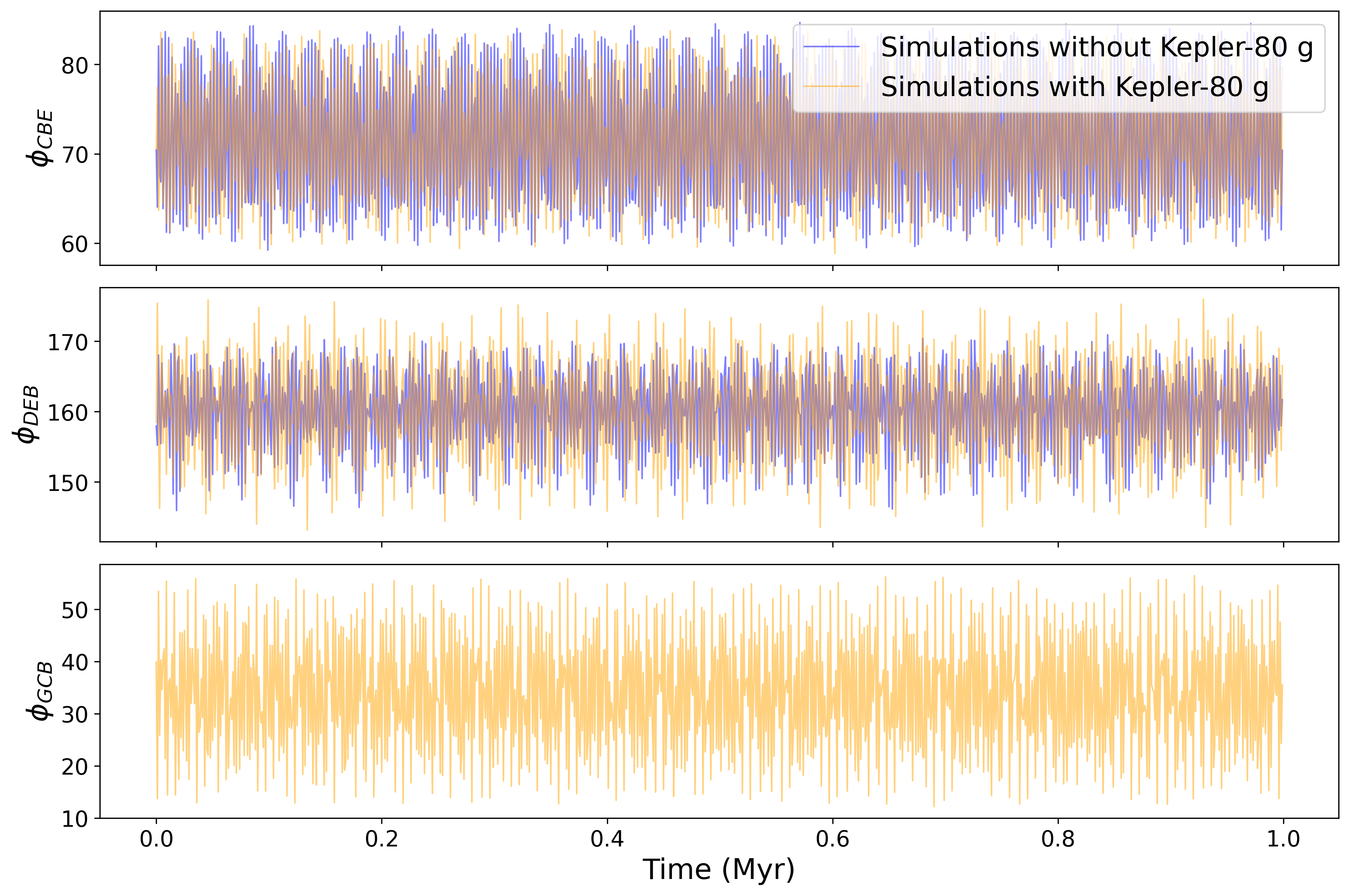}
    \caption{These plots show the {evolution} of the resonant angles {(given in degrees)} for {two draws from the unconstrained eccentricity posteriors (one with Kepler-80 g and one without Kepler-80 g) for a case where the planets are in a five-planet resonant chain. The evolution was computed by N-body simulation.} }
    \label{fig:rescurve}
\end{figure*}

To {assess the resonant state of the system,} we ran a large suite of numerical simulations using Rebound and the default IAS15 integrator \citep{Rein2012}. {For each of the four permutations of fit assumptions (e=0/floating eccentricity and including/excluding Kepler-80 g), we ran 500 simulations corresponding to 500 unique draws from the posteriors for each of the solutions presented in Table \ref{bigtable}. For all planets, the orbital inclinations were set to be coplanar (90 degrees), and the longitudes of ascending node set to 0 degrees. } 
{The integration length was chosen to minimize computation time but maximize coherence between the derived results and the results inferred from longer integrations. To choose the integration length, we ran a smaller subset of 35 integrations with lengths of 1 Myr. An example of the full 1 Myr evolution of the resonant angles is shown in Figure \ref{fig:rescurve}. For 33 of the 35 integrations, the results inferred from the first $10^{4}$ years of the integration were consistent with the full integration. Because of this, we adopt $10^{4}$ years as the integration length moving forward under the inference that $\sim95\%$ of the $10^{4}$ years integrations will yield the same results that a 1 Myr integration would have yielded. Since we are considering the behavior of the full ensemble of integrations, this is accurate enough to derive the general behavior of the libration. }

\begin{figure*}
    \centering
    \includegraphics[width=12cm]{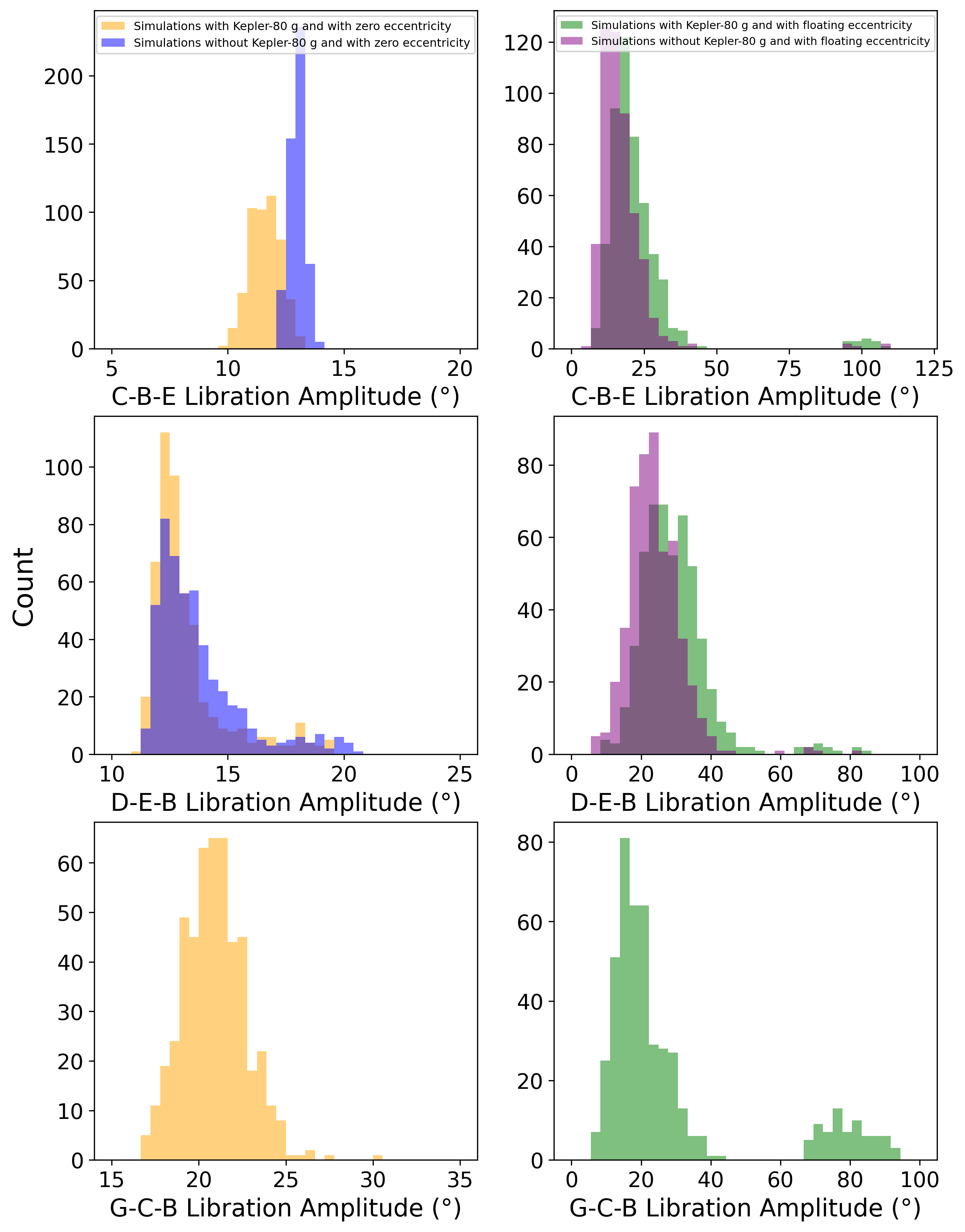}
    \caption{{These histograms show the simulated libration amplitudes for the Kepler-80 three-body MMRs. For each of the four fits given in Table \ref{bigtable}, 500 randomly selected draws were taken from the posterior and for each one, a numerical integration was run for 10$^{4}$ years. From these integrations, the libration amplitude was computed. The vast majority of the draws from the posteriors show resonant libration for all three three-planet chains considered. Note that 3.8\% of simulations from the fit including Kepler-80 g and with floating fit eccentricity exhibited non-resonant behavior in the g-c-b resonance, and were not included in the histograms shown here.}}
    \label{fig:reshist}
\end{figure*}

{For each integration, the initial conditions for the simulation were taken from the chosen draw from the posterior. From the results of the N-body simulations, we} computed the libration {amplitude} of {each} resonance. {The result of the 2000 total integrations (500 per posterior) is} shown in Figure \ref{fig:reshist}. 
The libration {amplitude} of the three-body {MMRs} were calculated by finding one-half the difference between the maximum and minimum angles formed by the three planets over the {10$^{4}$}-year {length of the integration}.

{Figure \ref{fig:reshist} demonstrates the differences in libration amplitude between the cases considered in this work. The draws from the fits where the orbital eccentricities were fixed to zero (left panels) have significantly smaller computed libration amplitudes than the draws from fits where eccentricity was unconstrained (right panels). In the latter case, a small fraction of draws result in libration amplitudes around a second center with a much higher amplitude, corresponding to a different position in dynamical phase space. (When this occurs, the high measured amplitude is due to the midpoint of the oscillations of the resonant angle changing throughout the numerical integration, rather than the oscillation having a large amplitude at any given time.) The draws from the $e=0$ solutions also have a lower overall range in potential libration amplitudes.}
{The difference of adding Kepler 80 g to the system is comparatively small, with only slight differences in the libration amplitude distributions between the two cases.}

{Most importantly, the fit with unconstrained eccentricities exhibit a wider range of dynamical behavior than the fit with $e=0$. While all of the draws from the $e=0$ posteriors were found to be fully resonant (in all three angles) for the length of the integrations, this was not the case in the unconstrained eccentricity fits.}
{In the majority of draws, we found the b-c-g angle to be librating throughout the simulation. However, in 3.8\% of draws from the posterior of the fit including Kepler-80 g and without constraints on planetary eccentricities, this angle exhibits nodding in and out of resonance. Because the angle librates for the full length of simulations in the vast majority of draws, this seems to indicate that Kepler-80 g is very likely a member of the resonant chain.}

{It is also important to note the difference between the results found here and the results found in \citet{MacDonald2021}. Most notably, while we found that Kepler-80 g is a member of a three-body resonance with members of the preexisting resonant chain in 96.2\% samples drawn, \citet{MacDonald2021} only found this to be true 8.2\% of the time. This is likely to due to the methodological differences previously discussed, especially the different fitting techniques of this paper and of \citet{MacDonald2021}. }

\subsection{Two-body resonances}
{T}wo-body {MMRs} exist when a configuration of planets has librating resonance angles, where the {component angles are} given by:
\begin{equation}
\phi_1 = p\lambda_i - q\lambda_j - (p-q)\omega_{i}, 
\label{eq:angle1}
\end{equation}
and 
\begin{equation}
\phi_2 = p\lambda_i - q\lambda_j - (p-q)\omega_{j}, 
\label{eq:angle2}
\end{equation}
where $\lambda$ is as given at the beginning of the section and the period ratio of the two planets is $p/q$ \citep[ex:][]{Wang2017}. 
{The composite two-body resonance angle, which reduces the full Hamiltonian to one degree of freedom \citep[as used in][]{Batygin2015,Petit2020, Goldberg2022}, is given by Equation 39 in \citep{Laune2022} to be}
\begin{equation}
\tan{\Phi_{12}} = \frac{f_{1}e_{1} \sin{\phi_1} + f_{2}e_{2} \sin{\phi_2}}{f_{1}e_{1} \cos{\phi_1} + f_{2}e_{2} \cos{\phi_2}}.
\label{eq:anglefull}
\end{equation}
{where coefficients $f_{1}$ and $f_{2}$ vary depending on the values of coefficients $p$ and $q$ (see Table 1 of \citealt{Deck2013}). This angle is called the "mixed" resonant angle in \citet{Goldberg2022}. In multi-planet systems, libration of the angles given in Equations \ref{eq:angle1} and \ref{eq:angle2} denotes a two-body mean motion resonance. However, as discussed in \citet{Goldberg2022}, planets can still be in two-body resonance even if the angles given by Equations \ref{eq:angle1} and \ref{eq:angle2} circulate, so long as the mixed angle still librates. An example of a multi-planet system where this occurs is given in \citet{Dai2022}, and the planets in this system appear to be in both three-body and two-body MMR despite Equations \ref{eq:angle1} and \ref{eq:angle2} not necessarily librating.}

{While systems in three-body resonance but not in two-body resonance are not without precedent \citep{Gozdziewski2016, Mills2016, Goldberg2021}, systems in three-body resonance but not two-body MMR indicate that some type of dissipative process that occurred after the planetary migration. It is valuable to classify the true resonant state of a system to determine where it fits into this paradigm. Motivated by the importance of such classification, we used our newly derived posteriors to assess the state of two-body MMRs in the system.}

{Kepler-80 has previously been characterized \citep{MacDonald2016, MacDonald2021} as a system with planets that reside in three-body MMRs but not two-body MMRs. Those works used the component angles Equations \ref{eq:angle1} and \ref{eq:angle2} to assess the state of two-body MMRs in the system.}
We first began by checking for the presence of any two-body MMRs observed in the system using the simulations generated in the previous section. 
{For the final fit posterior (the fit with unconstrained eccentricities and Kepler-80 g included), we took the subset of draws for which $10^4$ year integrations were run in Section \ref{sec:31} and computed the mixed resonant angle (Equation \ref{eq:anglefull} for each pair of planets. For completeness, we also computed the component angles given by Equations \ref{eq:angle1} and \ref{eq:angle2} for each planet pair.}

Consistent with the results from \citet{MacDonald2016}, we find that in general two-body MMRs do not exist between the original five planets of the Kepler-80 system {for either the component angles previously considered or for the mixed angle}. Rarely, our simulations showed evidence of a two body resonance between one pair of planets for some amount of time, but overall the simulations dominantly showed a lack of two-body MMRs.
{In addition, none of our simulations resulted in secure libration of any of the two-body resonance angles involving Kepler-80 g.}

These preliminary investigations confirm the result of \citet{MacDonald2016} that the Kepler-80 planets, while in stable three-body MMRs, {do not appear to be} in two-body resonance with their observed parameters. However, due to uncertainties in the TTV fit, it is possible that the phase space populated by the TTV solutions does not include the true orbits of the planets. {Additionally, the formula given in Equation \ref{eq:anglefull}, which we used to check the resonant state of the system, was derived from the partial two-planet Hamiltonian \citep[Equation 1 of][]{Goldberg2022}. An interesting avenue for future work would be to derive the full mixed angle for a five planet system and determine if that angle shows any evidence of libration for the Kepler-80 planets}.

 \subsection{{Comparison to the Results of \citet{MacDonald2021}}}
 {In the previous subsections, we have shown that the posteriors produced from the center-time-of-transit TTV fits in this paper suggest that the outer planets in the Kepler-80 system most likely form a five-planet resonant chain. We have also shown that are posteriors do not indicate that the planets are in two body residence. This result, produced by posteriors generated by an independent fit method, supports the previous claim by \citet{MacDonald2016} that Kepler-80 fits the archetype of a multi-planet system where the planets are in three-body MMRs but not two-body MMRs \citep[as is also the case in Kepler-60, Kepler-223 and Kepler-221][]{Gozdziewski2016, Mills2016, Goldberg2021}.}
 
{The photodynamical fit of \citet{MacDonald2021} also evaluates the same five planets considered in this work. In this work, we find that none of the planets are in secure two-body MMRs, even when the full mixed angle is considered. \citet{MacDonald2021} also finds that the planets are not in two-body MMRs a majority of the time. However, the major difference between the results of \citet{MacDonald2021} and those of this work are in the fraction of the time all five planets participate in the resonant chain: \citet{MacDonald2021} finds that this occurs 8.2\% of the time, as compared to the 96.2\% computed in this work.}

{The main reason for the difference in the fraction of time or five plants participate in the resonant chain between these two works is the measured orbital predators of Kepler-80 g. Kepler-80 g is found to have a moderate eccentricity of $e_g = 0.13^{+0.0034}_{-0.0037}$ in \citet{MacDonald2021}, while in this work we find it to have a much lower value of $e_g = 0.02^{+0.03}_{-0.02}$, and the masses are similarly discrepant. As a result, the two solutions live in different regions of parameter space, resulting in significantly different inferred rates of Kepler-80 g's participation in the resonant chain.}

\section{Discussion} \label{sec:disc}
In this work, we consider the dynamics of the Kepler-80 system. {The Kepler-80 system was previously considered in \citet{MacDonald2016, MacDonald2018, MacDonald2021}, and is remarkable because of its chain of three-body MMRs. The purpose of this work was to demonstrate the differences in final inferred orbital parameters that various assumptions can cause, as well as fit the data using a different method that has been used previously.}
{In particular, one assumption we were interested in testing was} how our understanding of {Kepler-80's} dynamical state changes with the addition of Kepler-80 g, a low signal-to-noise detection missed in the original analysis and later reported in \citet{Shallue2018}. It is important to know the limitations of physical parameters derived from the data, since it is possible that more systems than just Kepler-80 host undiscovered planets which may perturb their dynamical states. 

\subsection{Transit Timing Variations: Insights and Limitations}
We first derived planetary masses using the transit timing variations, using {four} different fits with different assumptions about the planets present in the system. 
{The solutions for each of these four permutations are presented in Table \ref{bigtable}.}
In the TTV fit assuming null eccentricities, we find a planetary mass for the new planet Kepler-80 g to {be $m_g = 0.8 \pm 0.3 M_{\oplus}$. When we allow the eccentricities to float, the derived mass was $m_g = 1.0 \pm 0.3 \ M_{\oplus}$.} {These values are both very different from the mass measured by the photodynamical fit of \citet{MacDonald2021}, which was $m_g = 0.065_{-0.038}^{+0.044}$.  Our hypothesis for the discrepancy between the results of this work and that of \citet{MacDonald2021} is that the latter was biased to higher eccentricities due to low-SNR data on the events of Kepler-80 g and the nature of photodynamical fits. Due to low-SNR data,  \citet{MacDonald2021} itself cautions readers not to draw conclusions from their measured mass of Kepler-80 g. The mass presented in this work, in contrast, is much less precise (the mass posterior shown in Figure \ref{fig:histogram_per_mass} subtends a huge range of possible masses) and less biased}. 

{Since the measured central values of masses derived via TTV methods can change depending on the quantity of planets included in the model, it is important to be aware that}
masses derived for systems where not all planets {have been discovered} may misrepresent the true masses of the planets. This may be particularly important in studies of planetary compositions, atmospheric mass loss, or tidal parameters. 



Quite often, observational data is not enough to securely classify the dynamics of an exoplanet system {via TTV fits \citep[ex:][]{Gillon2016, Gillon2017}, and improvements can only be made once additional data is added into the model \citep{Agol2021}.} 
{When data quality is high enough to resolve the synodic chopping signal \citep{Deck2015chop}, the mass-eccentricity degeneracy can be broken. This is more feasible for systems with large TTV amplitudes, such as KOI-142/Kepler-88 \citep{Weiss2020}.}
{In the case of Kepler-80, where the data for the outer planet is particularly low-SNR and the baseline of the observations limited, the TTV solution could be improved with additional} data over {extended baseline} to better identify and fit the chopping signal. It seems feasible for ground-based observations to provide this improved data: transits of both Kepler-80 b and c should be detectable with the Wide Field Infrared Camera (WIRC) at the Palomar Observatory, for example \citep{Vissapragada2020}. 
It is also important to note that full TTV solutions may only be possible when additional effects and parameters \citep[including, for example, the effect of general relativity and variations in stellar and planetary tidal parameters;][]{Bolmont2020, Gomes2021} are included, further increasing the amount of data needed for a robust solution.

\subsection{Results on Resonance}

For the Kepler-80 system in particular, the most notable dynamical property of the system is the multi-planet resonant chain.
Figures \ref{fig:reshist} and Figure \ref{fig:rescurve} show that while the {libration amplitude} of the resonant angle does change slightly when the additional planet is added to the system, the difference is relatively small and does not change whether the system is in three-body resonance. {Additionally, we find Kepler-80 g participates in the resonant chain with a high probability, as 96.2\% of our N-body integrations show a secure five-planet resonant chain.}

As discussed in the review by \citet{Raymond2020}, understanding the likelihood of resonance as a dynamical outcome is imperative to understanding the larger processes of planet formation and migration.
Resonant chains are thought to form during the early history of the system, when the gas disk was still present, and as a result their continued existence billions of years later provides stronger constraints on the intervening dynamics than can be found for most non-resonant systems. 
Precisely characterized resonant chains can provide a window to both their formation state and subsequent evolution \citep{Delisle2017, Goldberg2021}. 
In particular, the occurrence of librating three-body MMRs without simultaneous two-body MMRs \citep{Gozdziewski2016, Mills2016, MacDonald2016, Goldberg2021} provides unique constraints on the migration process. 

Assessing the presence of two-body and three-body MMRs in a system requires relatively precise characterization of the planets' orbital elements, and two-body MMRs in particular depend strongly on the planetary $e$ and $\omega$. Because of this, when three-body MMRs are found but two-body MMRs are not found, there are two credible causes: (1) a dynamical origin which can tell us more about the planetary migration history, or (2) errors or uncertainties on the observationally-determined orbital elements that prevent the two-body MMRs from being seen. 
The results of \citet{MacDonald2016}, \citet{MacDonald2021}, and this work all show that the middle four planets in the Kepler-80 system are in secure three-body MMRs but not two-body MMRs. As such, although it is possible that further data could refine the orbital solution into a parameter space where the planets are also in two-body resonance, {all interpretations of} the current observational evidence suggests that the Kepler-80 system does not exhibit two-body resonance. 

{The only point regarding resonance on which the results of this work disagree with previous work is on the participation of Kepler-80 g in the resonant chain.} \citet{MacDonald2021} found that Kepler-80 g could not be securely classified in the resonant chain {and} was a member of the chain with a 8.2\% probability, {as compared to our 96.2\% probability}. This conflict is due to the differing TTV fit methodologies and the substantial eccentricity for Kepler-80 g found in \citet{MacDonald2021}. More data on the system over a long baseline would help determine the true state of the system, but in either case the Kepler-80 system is a uniquely strong example of mean-motion resonance. 

Additional data would also help reconcile the differences between the solution of this work and the solution of \citet{MacDonald2021}. While both solutions are consistent with the data, they provide very different pictures of the system{, and as such emphasizes the importance of finding the mass and resonant state of planetary systems in multiple ways based on different methodologies for future work regarding systems where such properties are important.}

{\subsection{Secure Conclusions on Kepler-80}} 

{In this work, we used four permutations of center-time-of-transit fit assumptions to demonstrate the spread of solutions when different assumptions are used. In \citet{MacDonald2021}, the fit was performed using a photodynamical fit, a method subject to different biases than the method in this paper. Because of the differences between these various techniques, there is a opportunity to determine which conclusions about the Kepler-80 system are the most robust. Conclusions confirmed regardless of fit assumptions or methods are secure to use in dynamical analysis moving forwards.}

{The state of the system, which is multi planet system where planets are in three-body MMRs but not in two-body MMRs, is strengthened by the fact that all the results of this work and \citet{MacDonald2021} find the same result. While the two methods used different fit methodologies and compute  different posteriors, the ultimate dynamical conclusion made from those posteriors (that Kepler-80 is a system with three-body MMRs but not two-body MMRs) is the secure.}

{In comparison, the biggest difference between the results of this work and the results of \citet{MacDonald2021} are (i) the rate of Kepler-80 g's participation in the resonant chain, and (ii) the specific planetary parameters measured. The two methods have different biases, with the \citet{MacDonald2021} result being more susceptible to bias from low-SNR data and the results of this work not including information from the transit durations. Even within this work, the four permutations of fit assumptions give significantly different parameters depending on the assumptions used in the fit. Future dynamical analyses that depend sensitively on the results that differ between the two works must consider the full parameter space described by both solutions.}

\input{bigtable}

\medskip
{Acknowledgements.}We thank Max Goldberg, Konstantin Batygin, Mariah MacDonald, and Eric Agol for useful conversations. 
D.W. thanks the UROP program at the University of Michigan for providing this research opportunity.  J.C.B.~has been supported by the Heising-Simons \textit{51 Pegasi b} postdoctoral fellowship.

Software: pandas \citep{ mckinney-proc-scipy-2010}, IPython \citep{PER-GRA:2007}, matplotlib \citep{Hunter:2007}, scipy \citep{scipy}, numpy \citep{oliphant-2006-guide}, Jupyter \citep{Kluyver:2016aa}, TTVFast \citep{Deck2014soft}, Rebound \citep{Rein2012}, Reboundx \citep{Tamayo2020}

\bibliography{refs}

\end{document}

%% file: bigtable.tex
\begin{deluxetable*}{c c c c c c c}
\rotate
\tabletypesize{\footnotesize}
\tablewidth{0pt}
 \tablecaption{Results of Orbital Parameters Calculated in Fits}
 \tablehead{
 \colhead{Parameter} & \colhead{Kepler-80} & \colhead{Kepler-80 d} & \colhead{Kepler-80 e} & \colhead{Kepler-80 b}& \colhead{Kepler-80 c} & \colhead{Kepler-80 g} 
 }
 \startdata 
\multicolumn{6}{c}{\textit{Results of Circular Fit Without Kepler-80 g}}\\ 
 Solar Mass [$M_\odot$] & $0.73 \pm 0.03$ \\
 Mass [$M_E$] & & $4.8 \pm 0.4$ & $6.2 \pm 0.5$ & $4.7 \pm 0.4$ & $4.1_{-0.3}^{+0.4}$\\
 Orbital Period [days] & & $3.072086^{+0.000009}_{-0.000008}$ & $4.64554 \pm 0.00002$ & $7.05416 \pm 0.00003$ & $9.52329 \pm 0.00003$ \\
 Mean Anomaly [°] & & $95.8 \pm 0.1$ & $171.41 \pm 0.09$ & $289.01 \pm 0.04$ & $132.61 \pm 0.04$ \\
 \\
 \multicolumn{6}{c}{\textit{Results of Circular Fit Simulated With Kepler-80 g}}\\
 Solar Mass[$M_\odot$] & $0.73 \pm 0.03$ \\
 Mass [$M_E$] & & $4.1 \pm 0.4$ & $6.3 \pm 0.5$ & $4.5 \pm 0.4$ & $4.0 \pm 0.3$ & $0.8 \pm 0.3$ \\
 Orbital Period [days] & & $3.07208 \pm 0.000008$ & $4.64551 \pm 0.00002$ & $7.05420 \pm 0.00003$ & $9.52318 \pm 0.00005$ & $14.6482 \pm 0.0002$ \\
 Mean Anomaly [°] & & $96.0 \pm 0.15$ & $171.42 \pm 0.09$ & $289.01 \pm 0.04$ & $132.58 \pm 0.04$\\
 Time of Transit [BJD - 2454900] & & & & & & $70.262 \pm 0.004
 $$^1$ \\
 \\
 \multicolumn{6}{c}{\textit{Results of Floating-Eccentricity Fit Without Kepler-80 g}}\\
 Mass [$M_E$] & $0.73 \pm 0.03$ & $4.7 \pm 0.4$ & $2.4_{-0.3}^{+0.4}$ & $2.3_{-0.5}^{+0.6}$ & $3.1_{-0.6}^{+0.7}$ & \\
 Orbital Period [days] & & $3.07219_{-0.00002}^{+0.00003}$ & $4.64537 \pm 0.00009$ & $7.05325_{-0.00009}^{+0.00008}$ & $9.5233 \pm 0.0002$ & \\
 Time of Transit [BJD - 2454900] & & $70.096_{-0.006}^{+0.004}$ & $67.604_{-0.007}^{+0.005}$ & $67.151_{-0.009}^{+0.006}$ & $72.394_{-0.009}^{+0.007}$ & \\
 Orbital Eccentricity$^2$ & & $0.005_{-0.003}^{+0.004}$ & $0.007_{-0.004}^{+0.003}$ & $0.005_{-0.003}^{+0.005}$ & $0.012_{-0.005}^{+0.006}$ & \\
 Argument of Pericenter[$°$]$^2$ & & $249_{-91}^{+100}$ & $65_{-39}^{+31}$ & $323_{-56}^{+46}$ & $185_{-12}^{+17}$ &
 \\
 \\
 \multicolumn{6}{c}{\textit{Results of Floating-Eccentricity Fit With Kepler-80 g}}\\
 Mass [$M_E$] & $0.73 \pm 0.03$ & $4.1 \pm 0.4$ & $2.2 \pm 0.4$ & $2.4 \pm 0.6$ & $3.4_{-0.7}^{+0.9}$ & $1.0 \pm 0.3$\\
Orbital Period [days] & &  $3.07221 \pm 0.00003$ & $4.6453_{-0.00009}^{+0.00010}$ & $7.05325 \pm 0.00009$ & $9.5232 \pm 0.0002$ & $14.6471^{+0.0007}_{-0.0012}$\\
 Time of Transit [BJD - 2454900] & & $70.096_{-0.005}^{+0.003}$ & $67.606_{-0.006}^{+0.005}$ & $67.150_{-0.010}^{+0.007}$ & $72.392_{-0.010}^{+0.008}$ & $70.3 \pm 0.1$ \\
 Orbital Eccentricity$^2$ & & $0.005_{-0.003}^{+0.004}$ & $0.008 \pm 0.004$ & $0.006^{+0.005}_{-0.004}$ & $0.010^{+0.006}_{-0.005}$ & $0.02^{+0.03}_{-0.02}$\\
 Argument of Pericenter [°]$^2$ & & $260^{+98}_{-89}$ & $71^{+30}_{-27}$ & $326^{+39}_{-52}$ & $190^{+26}_{-16}$ & $153^{+72}_{-55}$
 \enddata
 \tablecomments{This table details the means and standard deviations of the calculated masses, periods, and mean anomalies for each of the five planets in optimization calculations, simulated with and without the existence of Kepler-80 g, the former being simulated with the eccentricity both being set to 0 and allowed to float. The solution was computed at epoch 2454964 BJD. Note that in the fit with free-floating eccentricity, the precision on all fitted parameters is worse than in the other two cases due to the increased number of fitted parameters.$^1$: {Since individual transits of Kepler-80 g were not modeled, th}e time of transit {for Kepler-80 g} was computed using the value of the y-intercept (for transit number 0) of the
linear fit to transit times used to compute the period for Kepler-80 g. $^2$: The solution shows some non-zero eccentricity for Kepler-80 e and g. $e$ and $\omega$ were not fit directly, but derived from fit parameters $\sqrt{e}\cos{\omega}$ and $\sqrt{e}\sin{\omega}$.}
\label{bigtable}
\end{deluxetable*}